\RequirePackage{fix-cm}

\documentclass[dvitex,twocolumn,epjc3]{svjour3}          

\RequirePackage[T1]{fontenc}

\smartqed  

\RequirePackage{graphicx}
\RequirePackage{flushend}
\RequirePackage[numbers,sort&compress]{natbib}
\RequirePackage[colorlinks,citecolor=blue,urlcolor=blue,linkcolor=blue]{hyperref}

\journalname{Eur. Phys. J. C}

\usepackage[T1]{fontenc}
\usepackage{amsmath,amsfonts,amssymb}
\usepackage[mathcal]{eucal}
\usepackage{mathrsfs}
\usepackage{latexsym}
\usepackage{bm}
\usepackage[all]{xy}
\usepackage{dsfont}

\usepackage{caption}
\usepackage{subcaption}

\newcommand{\sech}{\text{sech}}
\newcommand{\e}{\text{e}}

\begin{document}

\title{A smoothed string-like braneworld in six dimensions}


\author{J. C. B. Ara\'{u}jo\thanksref{e1}
        ,
        J. E. G. Silva\thanksref{e2}
        , 
        D. F. S. Veras\thanksref{e3}
        \and
        C. A. S. Almeida\thanksref{e4}
        }

\thankstext{e1}{e-mail: julio@fisica.ufc.br}
\thankstext{e2}{e-mail: euclides@fisica.ufc.br}
\thankstext{e3}{e-mail: franklin@fisica.ufc.br}
\thankstext{e4}{e-mail: carlos@fisica.ufc.br}

\institute{Departamento de F\'{i}sica - Universidade Federal do Cear\'{a} \\ C.P. 6030, 60455-760
Fortaleza-Cear\'{a}-Brazil}

\date{Received: date / Accepted: date}

\maketitle

\begin{abstract}
We propose here a static and axisymmetric braneworld in six dimensions as a string-like model extension. For a subtle warp function, this scenario provides near brane corrections. By varying the bulk cosmological constant, we obtain a source which passes through different phases. The solution is defined both for the interior as for the exterior of the string and satisfies the weak energy condition. Smooth gravitational massless mode is localized on the brane which core is displaced from the origin. In contrast to the thin string model, the massive solutions have high amplitude near the brane. By means of an analogue quantum potential analysis, we show that $s-$waves gravitational Kaluza-Klein modes are permissible as resonant states.
\end{abstract}


\section{Introduction}
\label{Sec-Introduction}

In the last decade, the extra dimension models turned to be a cornerstone of the high energy physics \cite{Randall:1999ee,Randall:1999vf}. In particular, the Randall-Sundrum (RS) model brought the idea of infinity extra dimension through a warped geometry \cite{Randall:1999vf}. Some authors enhanced the RS model results to six dimensions. Since the two dimensional transverse manifold has its own geometry, it leads to some nonexistent features in the RS models. Indeed, for an axisymmetric brane, the so-called string-like defect, the brane tension is related with the conical deficit angle of the transverse space \cite{Olasagasti:2000gx,Gherghetta:2000qi,Ponton:2000gi,Cohen:1999ia,Gregory:1999gv}. Furthermore, the Kaluza-Klein (KK) modes produce a small correction to the Newtonian potential \cite{Gherghetta:2000qi}.
The gauge bosons can be trapped to the brane by means of only the gravitational interaction \cite{Oda:2000zc}. Moreover, it is possible to localize the vector and spinor fields in the same geometry \cite{Liu:2007gk}.

In spite of the aforementioned results, the string-like models exhibit some issues about their sources. In fact, these branes can be realized as a stable solution of a gauge and scalar fields \cite{Cohen:1999ia,Gregory:1999gv}.
Nonetheless, the solution for the Einstein field equations for these models are still lacking. Amongst the string-like models, one which the width of the string vanishes and it is possible to concern with only the vacuum solution is the Gherghetta-Shaposhnikov (GS) model \cite{Gherghetta:2000qi}. However, the thin
string-like models do not satisfy the dominant energy condition \cite{Tinyakov:2001jt}. In order to overcome it, some authors solved numerically the equations for an Abelian vortex in six dimensions which exhibits smooth geometry and satisfies all the energy conditions \cite{Giovannini:2001hh}.
In the supersymmetric approach, a realistic and smooth cigar solution was also found by numerical means \cite{deCarlos:2003nq}.

Another issue of the thin string-like models, is that all the regularity conditions are not satisfied at the origin \cite{Tinyakov:2001jt}. However, some authors added a conical behavior near the origin and studied its consequences \cite{Kehagias:2004fb,Garriga:2004tq,Gogberashvili:2007gg,Papantonopoulos:2005ma}, whereas other tried to smooth this conical behavior \cite{Papantonopoulos:2005ma, Bostock:2003cv, Kofinas:2005py, Kanno:2004nr, Cline:2003ak,  Papantonopoulos:2007fk, Navarro:2003vw, Navarro:2004di, Silva:2011yk}.

In this article we explore some features of a smooth extension of the GS model built analytically with a conical behavior. Besides,
using a smoothed warp function, the metric satisfies all the regularity conditions.
We analyse the geometrical and physical properties of this model which yields an interior and exterior string-like solution, allowing near brane corrections to the GS model.
The source exhibits a conical behavior near the origin and the brane core is displaced from the origin. By studying the gravitational modes, we find a smoothed localized zero mode peaked around the shifted brane core and some smooth correction to the KK modes. Furthermore, the quantum analogue potential possesses an infinite well around the origin which barrier is strongly dependent on the bulk cosmological constant value.

This work is organized as follows: in Section \ref{Bulk geometry}, we review the main characteristics of the string-like branes and we present our extension. In Section \ref{Physical properties}, we study the properties of the source from the Einstein equations. The Section \ref{Gravity localization} is devoted to the study of the massless and massive gravitational modes upon  the geometry. Finally, in Section \ref{Conclusions and perspectives}, some final remarks and perspectives are outlined.


\section{Smoothed string-like geometry}
\label{Bulk geometry}

In this section, we review the main properties and models of the string-like branes and propose another complete solution.

Let $\mathcal{M}_{3}$ a $3$-brane and $\mathcal{M}_{2}$ a two dimensional Riemannian manifold. A string-like brane $\mathcal{M}_{3}$ is a Lorentzian manifold with axial symmetry about a manifold $\mathcal{M}_{2}$. Defining a gaussian radial coordinate $\rho\in [0,\infty)$ and an angular coordinate $\theta \in [0,2\pi]$, a well-known ansatz for a warped six-dimensional string-like braneworld $\mathcal{M}_{6}=\mathcal{M}_{p}\times\mathcal{M}_{2}$ is given by \cite{Olasagasti:2000gx,Gherghetta:2000qi,Ponton:2000gi,Cohen:1999ia,Gregory:1999gv,
Oda:2000zc}
\begin{equation}
\label{stringlikemetric}
\begin{split}
ds^{2} & = G_{AB}(x^{\zeta},\rho,\theta)dx^{A}dx^{B} \\
       & =  \sigma(\rho)\eta_{\mu\nu}dx^{\mu}dx^{\nu} + d\rho^{2} + \gamma(\rho)d\theta^{2},
\end{split}
\end{equation}
where $\sigma$, is the so-called warp function, $x^{\zeta}$ are on-brane coordinates, $(\rho,\theta)$ are coordinates of the transverse manifold and $\gamma$ is an angular factor. In order to guarantee a smooth $3$-brane at $\rho=0$, the metric components must to satisfy the usual regularity conditions \cite{Kofinas:2005py, Kanno:2004nr, Navarro:2003vw, Navarro:2004di}
\begin{eqnarray}
\label{regularityconditions}
\sigma(0) = 1 ,\hspace{1cm} & \sigma'(0) = 0,\\
\gamma(0) = 0 ,\hspace{1cm} & (\sqrt{\gamma(0)})'=1,
\end{eqnarray}
where the prime stands for the derivative $\partial_{\rho}$.

A special analytical string-like braneworld is the GS model for which \cite{Gherghetta:2000qi,Giovannini:2001hh}
\begin{equation}
\label{thinstringbrane}
\sigma(\rho) = \e^{-c \rho} \hspace{0.5cm} \text{and} \hspace{0.5cm} \gamma(\rho)=R_{0}^{2}\sigma(\rho),
\end{equation}
where $c\in \mathbb{R}$ is a constant which dimension is $[c]=L^{-1}$ and $R_{0}$ is an arbitrary length scale. It is worthwhile to say that the ansatz $(\ref{thinstringbrane})$ satisfies only the first regularity condition. Furthermore, the scalar curvature for this ansatz is
\begin{equation}
\label{thinstringcurvature}
R = - \frac{15}{2}c^{2}.
\end{equation}
Hence, this metric represents a $AdS_{6}$ manifold. The metric $(\ref{thinstringbrane})$ can be regarded as an exterior solution of the string-brane of width $\epsilon$. For $\epsilon \rightarrow 0$, the brane is infinitely thin. An awkward property of the thin string-like branes is that they do not satisfy the dominant energy condition, what turns this model a fairly exotic scenario \cite{Tinyakov:2001jt}. In order to overcome these issues, some authors derived numerically the geometry from an Abelian vortex \cite{Giovannini:2001hh} and in a supersymmetric approach \cite{deCarlos:2003nq}. Here, we propose an extension of the thin string-like solution and we investigate the modifications on the geometrical and physical properties of this scenario.

In this work, we consider the following warp factor \cite{Silva:2012yj}
\begin{equation}
\label{warpfactor}
\sigma(\rho) = \e^{-(c\rho - \tanh{c\rho})}.
\end{equation}

\begin{figure}[!htb] 
        \begin{minipage}[b]{ \linewidth}
            \includegraphics[width=\linewidth]{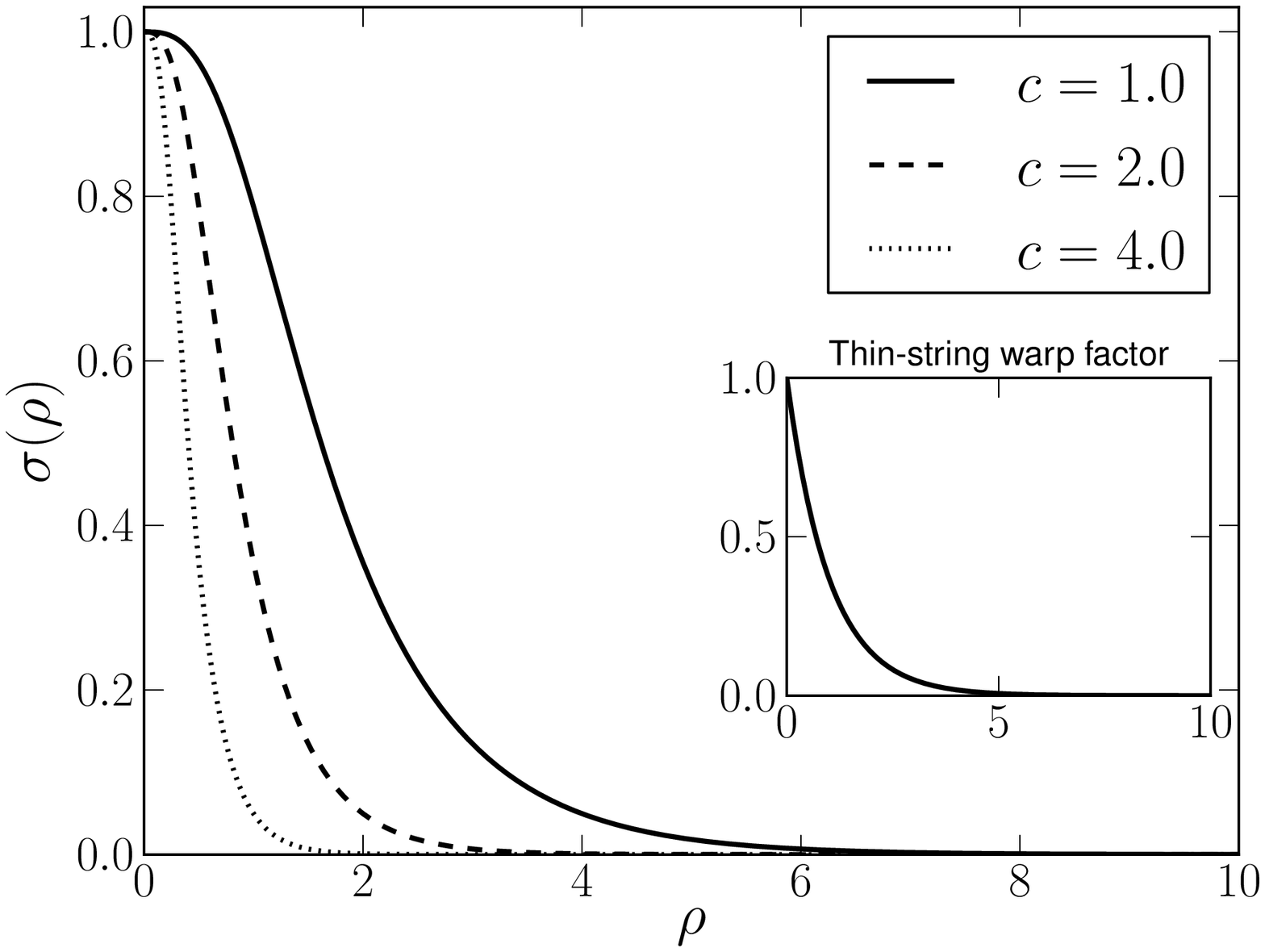}
            \caption{Warp function for some values of $c$. The thin string warp factor (subgraph) given by Eq. (\ref{thinstringbrane}) is defined only for the exterior of the string.}
            \label{warpfunction}
        \end{minipage}\hspace{0.5cm} \\[0.4cm]   
        \begin{minipage}[b]{\linewidth}
            \includegraphics[width=\linewidth]{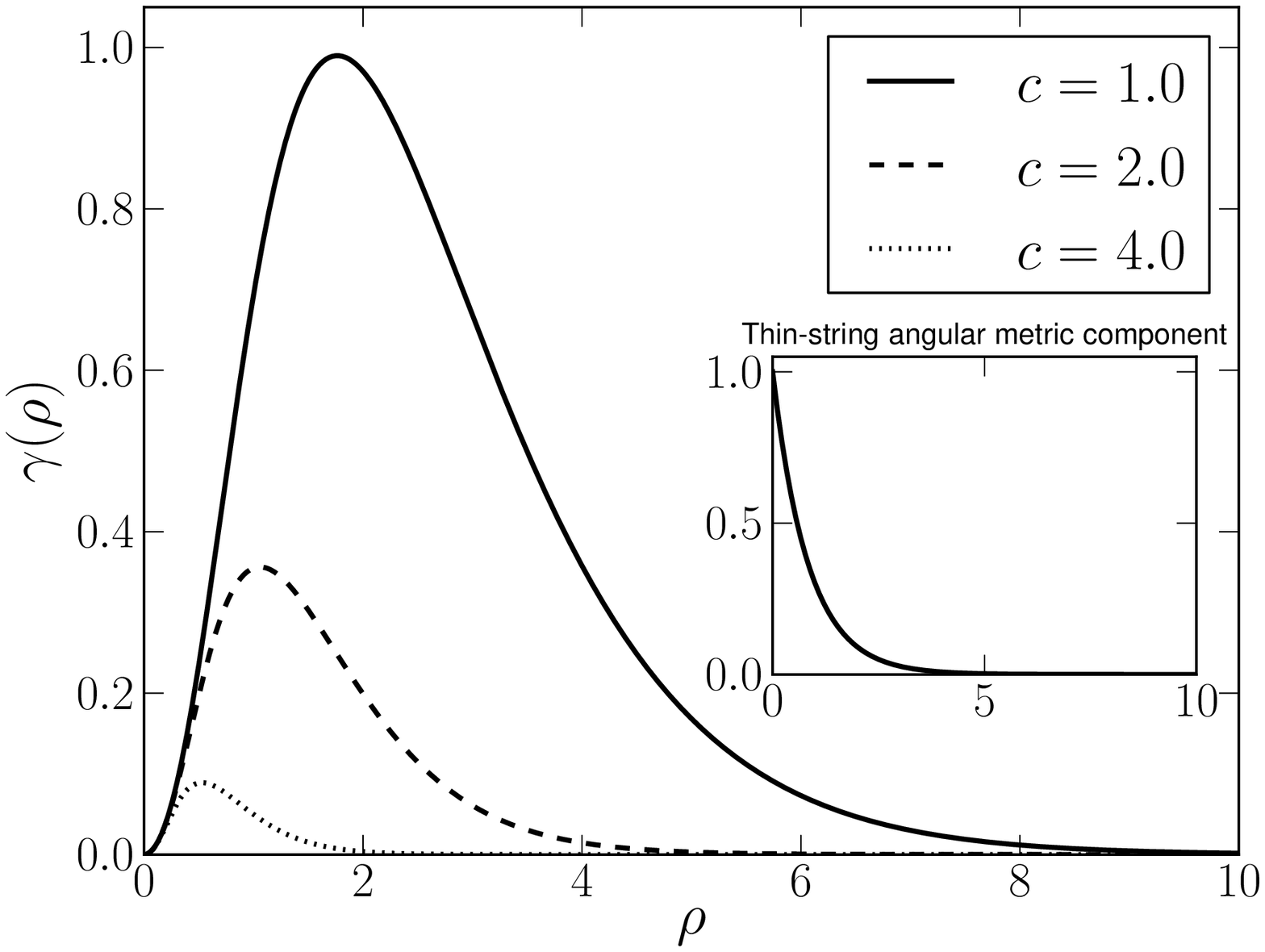}
            \caption{Angular metric component for different values of $c$. The subgraph refers to the thin string angular metric component  $(\ref{thinstringbrane})$ for $R_0 = 1.0$.}
            \label{gamma}
        \end{minipage}
\end{figure}

It is noteworthy to mention two important features of this warp function. First, as in the usual string-like geometries, it vanishes asymptotically \cite{Olasagasti:2000gx, Gherghetta:2000qi, Cohen:1999ia, Gregory:1999gv, Oda:2000zc, Tinyakov:2001jt, Giovannini:2001hh}. Second, unlike the thin string-like geometries \cite{Gherghetta:2000qi,Oda:2000zc}, it satisfies the regularity conditions. This last feature is due to the addition of the term $\tanh{c \rho}$ that smoothes the warp factor near the origin and it converges to the string-like one for large $\rho$ \cite{Silva:2012yj}. The warp function has a bell-shape as sketched in the Fig. \ref{warpfunction}. This agrees with the numerical solution generated by an Abelian vortex \cite{Giovannini:2001hh}. Therefore, we can realize this warp function as a near brane correction to the thin string-like models \cite{Olasagasti:2000gx, Gherghetta:2000qi, Oda:2000zc, Liu:2007gk}.

For the angular metric component, we have chosen the following ansatz:
\begin{equation}
\label{angularmetric}
\gamma(\rho) = \sigma(\rho)\beta(\rho),
\end{equation}
where $\beta(\rho)=\rho^{2}$. Note that all the string-like regularity conditions are now satisfied. We have plotted the angular component  $(\ref{angularmetric})$ in Fig. \ref{gamma}, where we show that the function exhibits a conical behavior near the brane. The $\rho^2$ term has strong influence near the origin. Moreover, both the warp function and angular component, have a $Z_{2}$ symmetry about the origin acquiring a bell-shape. This feature is not present in numerical solution for the Abelian vortex model \cite{Giovannini:2001hh}.


\begin{figure}[h]
\centering
\includegraphics[width=\linewidth]{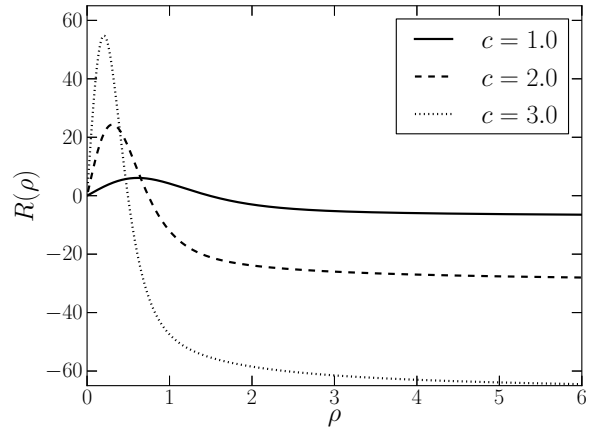}
\caption{Bulk scalar curvature.}
\label{Fig-ScalarCurvature}
\end{figure}

The scalar curvature of $\mathcal{M}_{6}$ is
\begin{equation}
\begin{split}
 R = c^2 & \left(10 \tanh{(c \rho)} \text{sech}^2{(c \rho)} - \frac{15}{2} \tanh^4{(c \rho)}\right) + \\ 
         & + 6 c \frac{\tanh^2{(c \rho)}}{\rho}.
\end{split}
\end{equation}
It was plotted in Fig. \ref{Fig-ScalarCurvature}. The curvature is everywhere smooth and it is noteworthy that $\mathcal{M}_{6}$ approaches to a $AdS_{6}$ asymptotically. This ensure us to claim that $(\mathcal{M}_{6},ds^{2}_{6})$ is an extension of the thin string models near and far the defect.

\section{Sources properties}
\label{Physical properties}
In this section, we study the components of the energy-momentum tensor as well as the value of the cosmological constant. 

The action for the gravitational field is defined as
\begin{equation}
\label{action}
  S_{g} =\int_{\mathcal{M}_{6}}{\left(\frac{1}{2\kappa_{6}}R-\Lambda +\mathcal{L}_{m}\right)\sqrt{-g}d^{6}x},
\end{equation}
where $\kappa_{6}=8\pi/M_{6}^{4}$, $M_{6}^{4}$ is the six-dimensional bulk Planck mass and $\mathcal{L}_{m}$ is the matter Lagrangian for the source of the geometry. Note that in this convention, the bulk cosmological constant $\Lambda$
has dimension
$[\Lambda]=L^{-6}=M^{6}$.

Consider an axisymmetric ansatz for the energy-momentum tensor as \cite{Gherghetta:2000qi,Oda:2000zc}
\begin{align}
\label{energymomentumansatz}
 T^{\mu}_{\nu} & = t_{0}(\rho)\delta^{\mu}_{\nu},\\
 T^{\rho}_{\rho} & = t_{\rho}(\rho),\\
 T^{\theta}_{\theta} & = t_{\theta}(\rho),
\end{align}
where
\begin{equation}
T_{AB}=\frac{2}{\sqrt{-g}}\frac{\partial (\sqrt{-g}\mathcal{L}_{m})}{\partial g^{AB}}.
\end{equation}\\
From the action (\ref{action}), we obtain the Einstein equation
 \begin{equation}
\label{Einstein}
 R_{AB}-\frac{R}{2}g_{AB} = -\kappa_{6}(\Lambda g_{AB} + T_{AB}).
\end{equation}
Using the metric ansatz (\ref{stringlikemetric}), the Einstein equation (\ref{Einstein}) yields to the components of the energy-momentum tensor in the form
\begin{eqnarray}
\label{energymomentumtensor}
t_{\theta}(\rho) & = & \frac{5c^2}{\kappa_{6}}\left[\text{sech}^{2}c\rho\left( 1 + \frac{4}{5} \tanh{c\rho}\right) -\frac{\text{sech}^{4} c \rho}{2}  \right],\\
t_{\rho}(\rho) & = & \frac{5c^2}{\kappa_{6}}\left(\text{sech}^{2}c\rho + \frac{2}{5}\frac{\tanh^{2}{c\rho}}{c\rho} -\frac{\text{sech}^{4} c \rho}{2} \right),\\
t_{0}(\rho) & = & t_{\theta}(\rho) + \frac{5c}{2\kappa_6}  \frac{\tanh^{2}{c \rho}}{\rho} .
\end{eqnarray}

\begin{figure*}
        \centering
        \begin{subfigure}[b]{0.49\linewidth}
                \includegraphics[width=\linewidth]{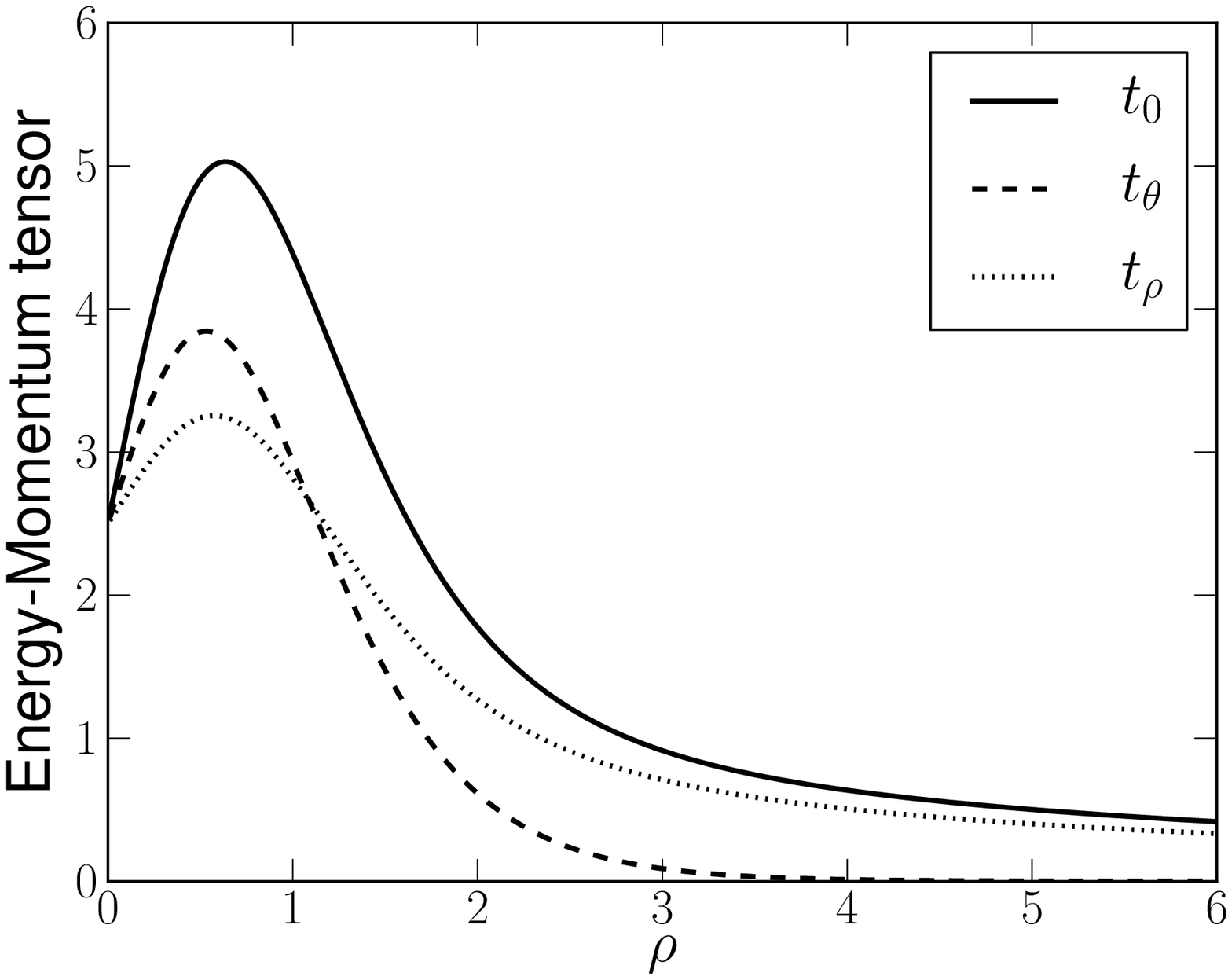}
                \caption{}
                \label{Fig-EnergyMomentum-1}
        \end{subfigure}%
        ~ 
        \begin{subfigure}[b]{0.49\linewidth}
                \includegraphics[width=\linewidth]{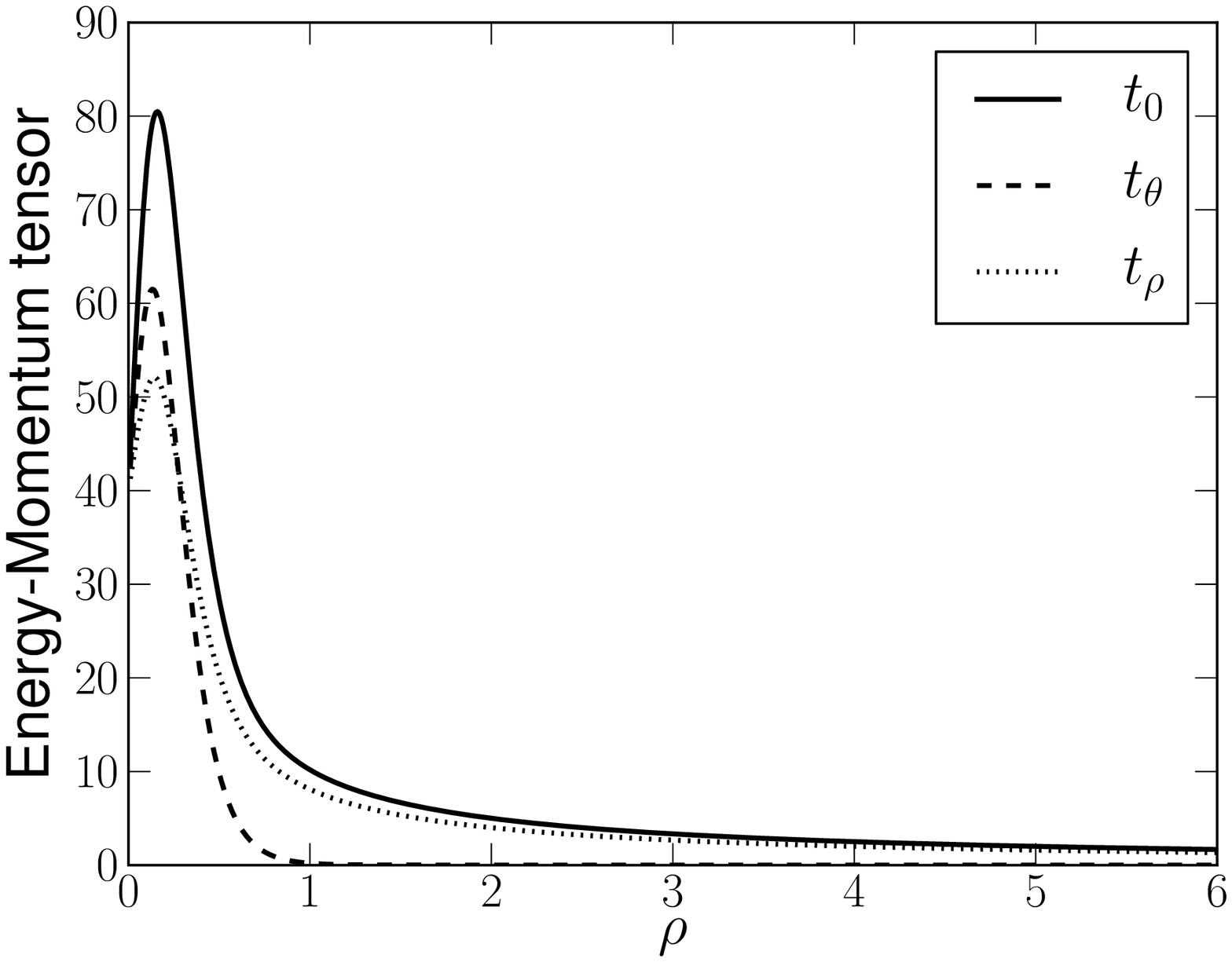}
                \caption{}
                \label{Fig-EnergyMomentum-4}
        \end{subfigure}
         ~

        \caption{Components of energy-momentum tensor for $c = 1.0$ (a)  and $c=4.0$ (b).}
         \label{Fig-EnergyMomentum}
\end{figure*}


For large $\rho$, the components of energy-momentum tensor vanish and the vacuum
solution of the Einstein equation yields the well-known relationship \cite{Gherghetta:2000qi,Oda:2000zc}
\begin{equation}
 c^{2}=-\frac{2}{5}\kappa_{6}\Lambda.
 \label{cosmologicalconstantrelation}
\end{equation}

The Eq. (\ref{cosmologicalconstantrelation}) determines the bulk to be asymptotically a $AdS_{6}$ spacetime. Besides, it also means that by varying $c$ we can study the changes in the source and fields for different bulk cosmological constant values.

We have plotted  in Fig. \ref{Fig-EnergyMomentum} the components of the energy-momentum tensor for $\kappa_6 = 1$. This smooth conical correction leads to a displacement of the core of the source from the origin. Similar results have been obtained by Giovannini \textit{et al} numerically for higher winding number Abelian vortex \cite{Giovannini:2001hh} and for the string-cigar model \cite{Silva:2012yj}. Moreover, the brane width defined as $\epsilon \approx \bar{\rho} - \rho_{\text{max}}$, where $\bar{\rho}$ and $\rho_{\text{max}}$  are the positions of the half-maximum and the maximum of $t_0$, respectively, tends to zero when $c\rightarrow \infty$. So, the GS model can be seen as a $c\rightarrow\infty$-limit of this smoothed version.


\section{Gravity localization}
\label{Gravity localization}

We now study the gravity localization on a $3-$brane embedded on this $6-$dimensional geometry. We consider a small perturbation $h_{\mu\nu}$ in the background metric in the form
\begin{equation}
ds^{2}_{6}=\sigma(\rho,c)(\eta_{\mu\nu}+h_{\mu\nu})dx^{\mu}dx^{\nu}+d\rho^{2}+\gamma(\rho,c) d\theta^{2}.
\end{equation}

Imposing the transverse traceless gauge $\nabla^{\mu} h_{\mu\nu}$ = $0$, the linearization of the Einstein equations yields to the equation for the gravitational perturbation \cite{Gherghetta:2000qi,Silva:2011yk}
\begin{equation}
\partial_{A} (\sqrt{-g_{6}} g^{AB} \partial_{B} h_{\mu\nu} ) = 0.
\end{equation}
Performing a well-known Kaluza-Klein decomposition \cite{Gherghetta:2000qi}
\begin{equation}
h_{\mu\nu} (x^{\zeta}, \rho , \theta) = \tilde{h}_{\mu\nu}(x^{\zeta})\sum_{m,l=0}^{\infty} \phi_{m,l}(\rho) \e^{i l \theta},
\end{equation}
where $l$ is an integer and $0\leq\theta\leq 2\pi$, and imposing the mass condition
\begin{equation}
\square_{4} \tilde{h}_{\mu\nu} (x^{\zeta}) = m^{2} \tilde{h}_{\mu\nu} (x^{\zeta}),
\end{equation}
the radial modes satisfy the Sturm-Liouville equation
\begin{equation}
\begin{split}
\frac{d^{2}}{d\rho^{2}} \phi(\rho) & + \left(\frac{1}{\rho} - \frac{5}{2}c\tanh^{2} (c\rho)\right) \frac{d}{d\rho}\phi(\rho) + \\ 
                                   & + M^{2}(\rho) \e^{c \rho - \tanh c \rho} \phi(\rho) = 0.
\end{split}
\label{Sturm-Liouville-Eq}
\end{equation}

The factor $M^{2}(\rho)$ = $\left(m^{2} - \frac{l^{2}}{\rho^{2}}\right)$ is an effective mass containing orbital angular momentum contributions $l$ and it behaves as a position dependent mass. Note that we have a degenerate KK spectrum, as in the GS model \cite{Gherghetta:2000qi}. However, since $\lim_{\rho\rightarrow\infty}{M^{2}(\rho)}=m^{2}$, the conical correction here breaks the degeneracy for large distances.

Due to the axial symmetry we impose the Neumann boundary conditions on $\phi_{m}$ \cite{Gherghetta:2000qi,Silva:2012yj}
\begin{equation}
\phi'_{m}(0) = \phi'_{m} (\infty) = 0.
\label{Neumann-Bound-Cond}
\end{equation}
Furthermore, these modes satisfy the following orthonormality condition:
\begin{equation}
\int_{0}^{\infty} \sigma(\rho) \sqrt{\gamma(\rho)} \phi^{*}_{m} \phi_{n} d\rho = \delta_{mn}.
\label{Ortonormalidade}
\end{equation}


\subsection{Massless mode}
\label{Sec-MasslessMode}

For $M^{2}(\rho) = 0$ (gravitational massless mode), a solution for the Eq. (\ref{Sturm-Liouville-Eq}) satisfying the boundary conditions (\ref{Neumann-Bound-Cond}) is the constant $\phi(\rho) = \phi_{0}$. However, from  the orthornormality relation (\ref{Ortonormalidade}), we obtain a varying zero-mode function $\psi_{0}$ as \cite{Gherghetta:2000qi,Silva:2011yk,Silva:2012yj}
\begin{equation}
\psi_0(\rho) = \phi_0 \sigma(\rho)\beta^{\frac{1}{4}}(\rho),
\label{ModoZero}
\end{equation}
where $\phi_0$ becomes the role of a normalization constant given by
\begin{equation}
\phi_0^2 =  \frac{1}{\int_0^{\infty} \sigma^{\frac{7}{2}}(\rho)\beta(\rho)d\rho}.
\end{equation}
In the next section we shall prove that the $\psi_{0}$ is the massless mode using the Schr\"{o}dinger approach. 

\begin{figure}
\centering
\includegraphics[width=\linewidth]{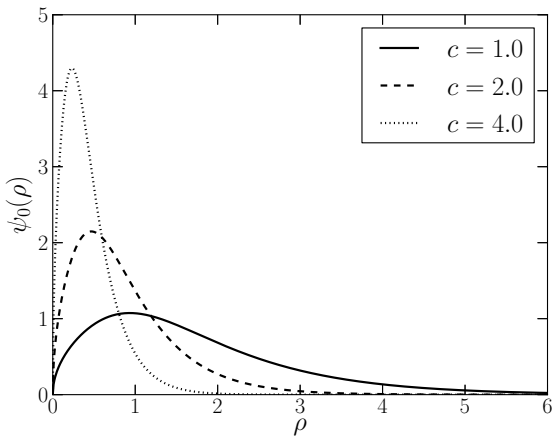}
\caption{Gravitational massless mode.}
\label{Fig-MasslessMode}
\end{figure}

Analysis of the massless mode graph, shown on Fig. \ref{Fig-MasslessMode}, reveal new results. Differently from thin-strings models, the maximum of $\psi_0$ is displaced from origin. It agrees with the fact that the brane core is not at $\rho = 0$. As showed in Section \ref{Physical properties}, the more the value of c more the brane core approaches to the origin, and this directly reflects on the zero-mode. This effect is due to the conical behavior near the origin induced by the $\beta$ factor. This mode goes along with the energy-momentum tensor. As $c$ increases, the maximum of the zero-mode and the energy density tend to coincide.  

Therefore, our smoothed model provides a near brane correction to the massless gravitational mode, smoothes the zero mode in the boundary of the core and recovers the thin-string exponential behavior for large distances.

\subsection{Massive modes}
\label{Sec-MassiveModes}

Solution of Eq. (\ref{Sturm-Liouville-Eq}) for $M^2(\rho) \neq 0$, called massive modes, are difficult to be obtained analytically. Although, before a numerical analysis, two different regimes are studied. 

Firstly, we  analyse graviton massive modes near the brane ($\rho\rightarrow0$). At this limit, the differential equation (\ref{Sturm-Liouville-Eq}) is reduced to a Bessel equation
\begin{equation}
\frac{d^{2}}{d \rho^{2}} \phi(\rho) + \frac{1}{\rho} \frac{d}{d \rho} \phi(\rho) + M^{2}(\rho)\phi(\rho) = 0,
\label{Eq-Bessel}
\end{equation}
which solutions satisfying the boundary conditions (\ref{Neumann-Bound-Cond}) are given by the Bessel functions of first kind $J_{l}(m \rho)$. Furthermore, the boundary condition (\ref{Neumann-Bound-Cond}) excludes the $l = 1$ solution, then, $p-$waves are not allowed for this model.

Asymptotically ($\rho\rightarrow\infty$), equation (\ref{Sturm-Liouville-Eq}) becomes
\begin{equation}
\frac{d^{2}}{d \rho^{2}} \phi(\rho) - \frac{5 c}{2} \frac{d}{d \rho} \phi(\rho) + m'^{2} \e^{c \rho} \phi(\rho) = 0,
\end{equation}
with a non-degenerate re-scaled mass  $m'$ = $m/\sqrt{\e}$. This equation is exactly the same as GS model, which solution is \cite{Gherghetta:2000qi}:
\begin{equation}
\begin{split}
\phi_{m^{\prime}} (\rho) =  \e^{(5/4) c \rho} & \left[  C_{1} J_{5/2} \left(\frac{2 m^{\prime}}{c} \e^{(c/2) \rho} \right) +  \right.\\
                                              & \left. + C_{2} Y_{5/2} \left(\frac{2 m^{\prime}}{c} \e^{(c/2) \rho} \right)\right],
\end{split}
\label{GS-MassiveModes}
\end{equation}
where $C_{1}$ and $C_{2}$ are arbitrary constants and $Y$ is the Bessel function of second kind.

\begin{figure*}
        \centering
        \begin{subfigure}[b]{0.49\linewidth}
                \includegraphics[width=\linewidth]{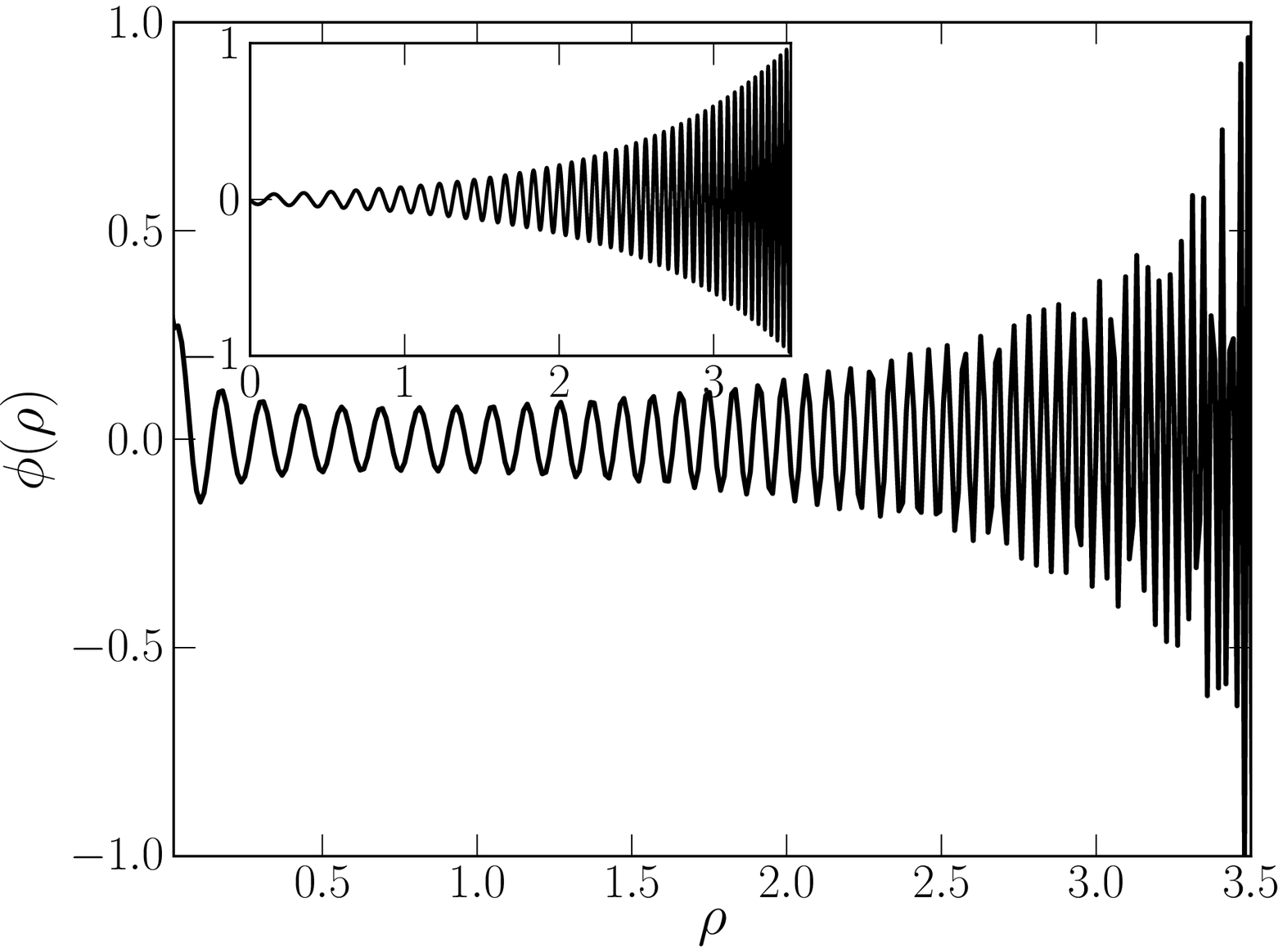}
                \caption{}
                \label{Fig-SolucaoNumerica1}
        \end{subfigure}%
        ~ 
        \begin{subfigure}[b]{0.49\linewidth}
                \includegraphics[width=\linewidth]{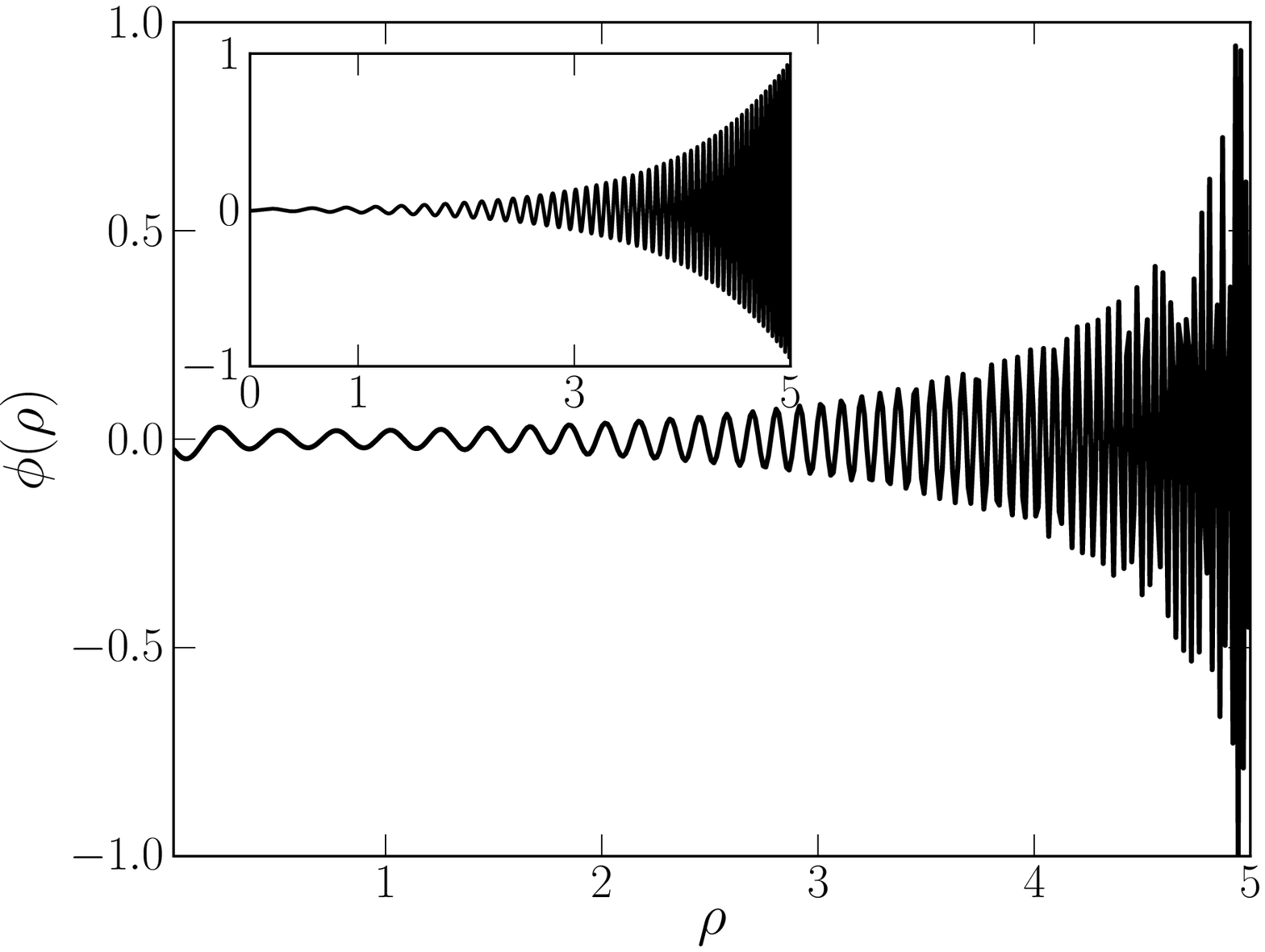}
                \caption{}
                \label{Fig-SolucaoNumerica2}
        \end{subfigure}
        ~ 

        \caption{Numerical solutions of the equation (\ref{Sturm-Liouville-Eq}) for $m = 44.10$ (a) and $m = 22.82$ (b). The sub-graphs are plots of the GS massive mode (\ref{GS-MassiveModes}) for the same mass values, where $C_1 = C_2 = 0.2$ (a) and  $C_1 = C_2 = 0.01$ (b) .}
         \label{Fig-SolucaoNumerica}
\end{figure*}

In order to obtain a complete domain solution, we solved Eq. (\ref{Sturm-Liouville-Eq}) numerically. For this purpose, we used the matrix method \cite{MatrixMethod} with second order truncation error for the domain $\rho \in [0,6]$. We have plotted in Fig. \ref{Fig-SolucaoNumerica} numerical solutions for $c = 1.0$, comparing with the analytical GS model solution (\ref{GS-MassiveModes}). Near the origin, the solutions behave as Bessel function, whereas far from the brane, as the GS massive modes solutions. The main advantage of the numerical task is that we obtain the full domain solution, so we may construe \textit{inside brane} massive gravitons. In contrast to the thin string model, the massive solutions are non-zero with high amplitude near the brane.

\subsection{Analogue quantum potential}
\label{Sec-QuantumPotential}

Another way to study massive modes is turn the KK equation (\ref{Sturm-Liouville-Eq}) into a Schr\"{o}dinger-like equation. This formalism provides knowledge about massive states that interact with the brane. Firstly, we perform a change of the independent variable as \cite{Silva:2011yk,Silva:2012yj}
\begin{equation}
z(\rho) = \int^{\rho} d\rho' \sigma^{-1/2},
\end{equation}
which provides a conformally plane metric and we write $\phi(z)$ as
\begin{equation}
\label{psifunction}
\phi_m(z) = \frac{1}{\sigma(z)\beta^{\frac{1}{4}}(z)} \psi_m(z),
\end{equation}
what yields the Eq. (\ref{Sturm-Liouville-Eq}) to a Schr\"{o}dinger equation for the $\psi_m(z)$ function
\begin{equation}
-\ddot{\psi_m}(z) + U(z) \psi_m(z) = m^{2} \psi_m(z),
\end{equation}
where the dots represents derivatives with respect to $z$ coordinate. The potential function has the form
\begin{equation}
U(z) = \frac{\ddot{\sigma}}{\sigma} + \frac{1}{2} \frac{\dot{\sigma}}{\sigma} \frac{\dot{\beta}}{\beta} - \frac{3}{16} \left(\frac{\dot{\beta}}{\beta}\right)^{2} + \frac{1}{4} \frac{\ddot{\beta}}{\beta} + \frac{l^{2}}{\beta}.
\end{equation}
From Eq. (\ref{psifunction}), we conclude the massless mode defined in Eq. (\ref{ModoZero}) satisfying the analogue Schr\"{o}dinger equation for $m = 0$. The potential can be rewritten on $\rho$ coordinate as
\begin{equation}
\begin{split}
\bar{U}(\rho;c,l) & = \e^{-(c \rho - \tanh c \rho)} \left( - 2 c^{2} \tanh (c\rho) \sech^2(c\rho)  \right. \\
                  & + \frac{3}{2} c^{2} \tanh^{4}(c\rho) - \frac{5c}{4}\frac{\tanh^{2}(c \rho)}{\rho} - \frac{1}{4 \rho^{2}} \\
                  & \left. + \frac{l^{2}\e^{(c \rho - \tanh c \rho)}}{\rho^{2}}  \right) .
\end{split}
\label{Potencial}
\end{equation}

\begin{figure*}
        \centering
        \begin{subfigure}[b]{0.49\linewidth}
                \includegraphics[width=\linewidth]{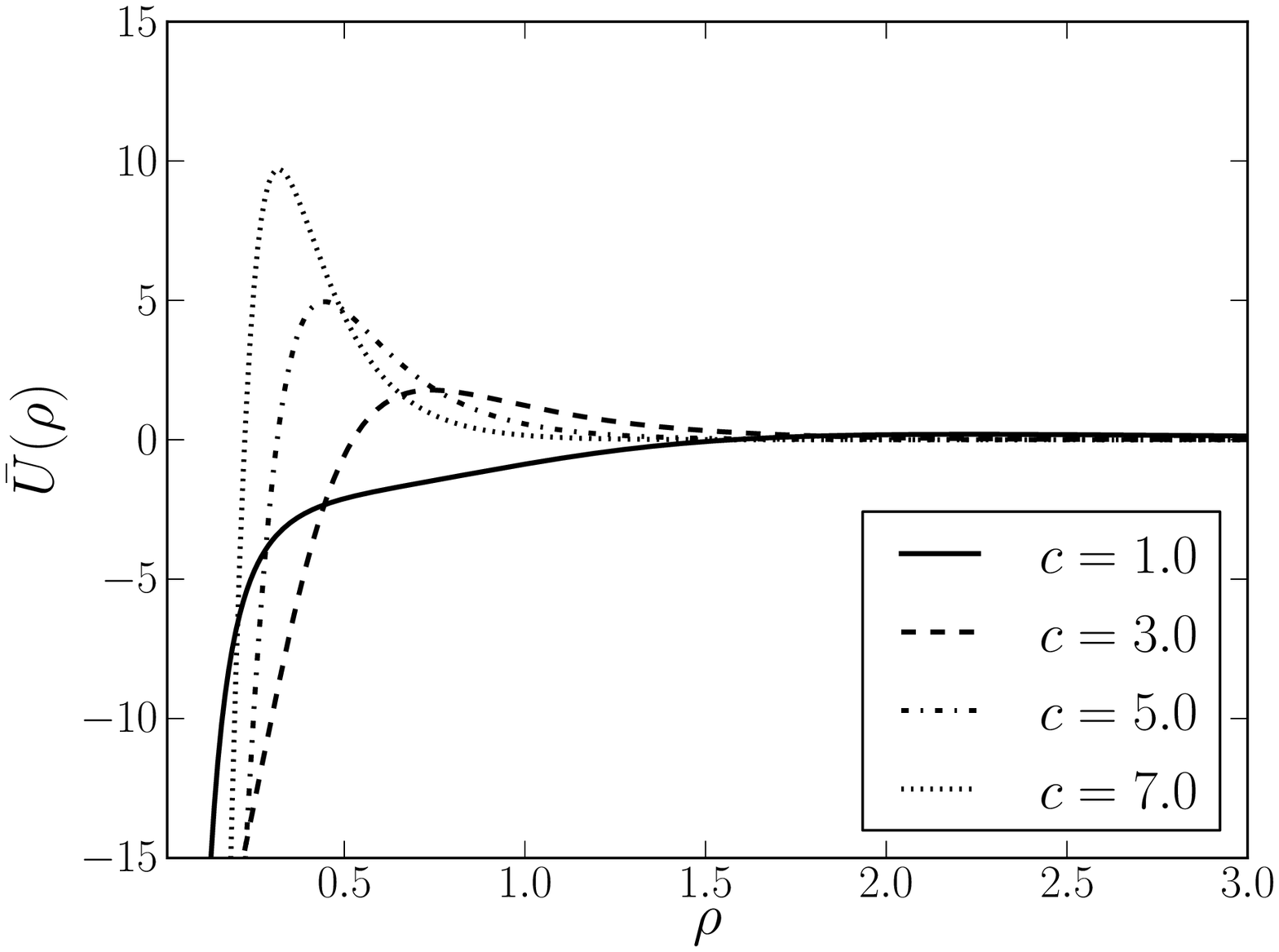}
                \caption{}
                \label{Fig-Potencial-0}
        \end{subfigure}%
        ~ 
        \begin{subfigure}[b]{0.49\linewidth}
                \includegraphics[width=\linewidth]{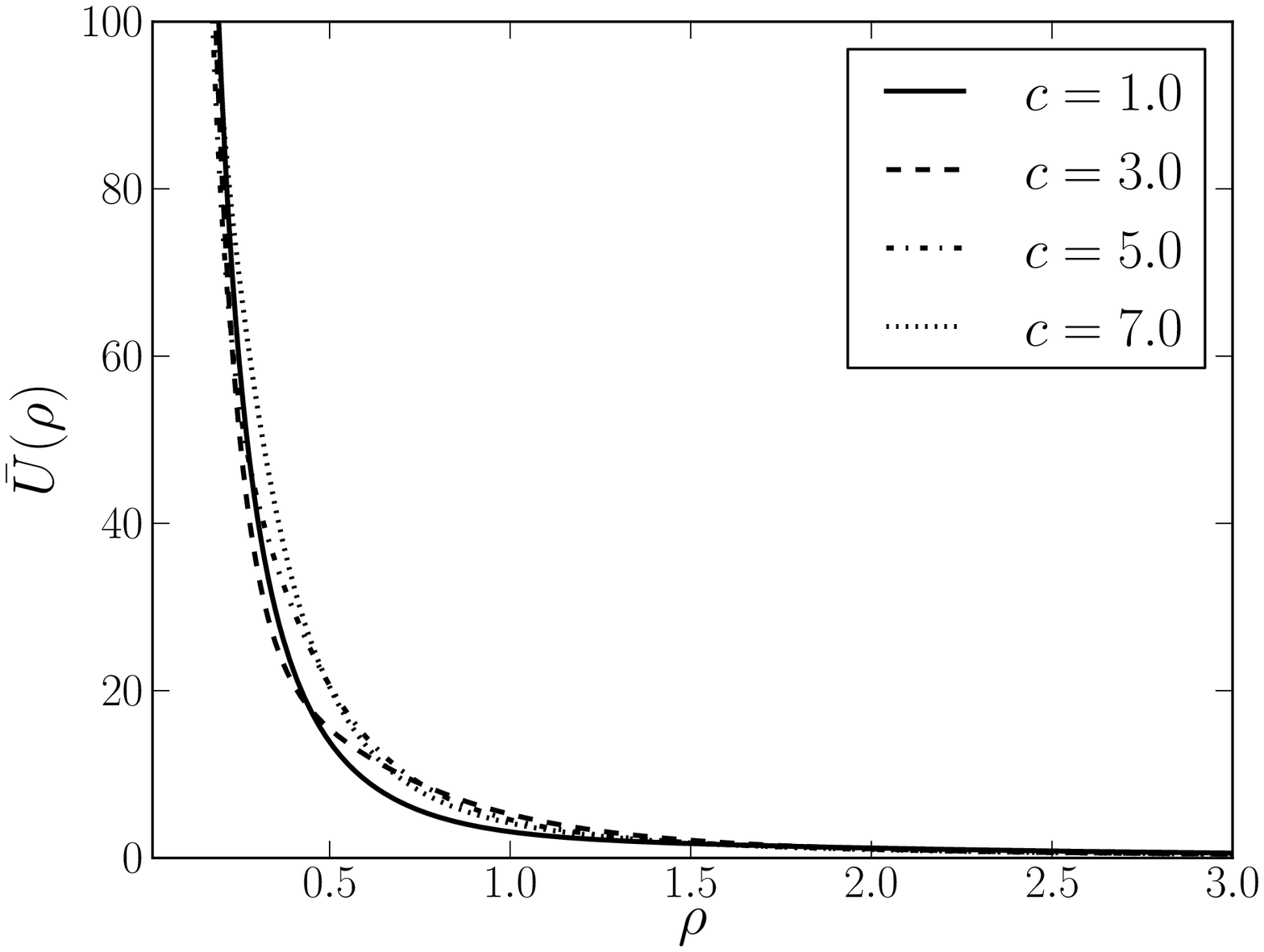}
                \caption{}
                \label{Fig-Potencial-1}
        \end{subfigure}
        ~ 

        \caption{Potential function $\bar{U}(\rho)$ for the Schr\"{o}dinger-like equation with $l = 0$ (a) and $l = 2$ (b).}
         \label{Fig-Potencial}
\end{figure*}

We have plotted the potential $\bar{U}(\rho)$ given by Eq. (\ref{Potencial}), on Fig. \ref{Fig-Potencial}. For $l = 0$ we have an infinite potential well at the origin arising the possibility of bound states where massive gravitons may interact with the brane as resonant states \cite{Ress-CASA, Ress-Wilami}. As $c$ increases, a barrier is formed besides the origin. On the other hand, for $l>1$, there is an infinite barrier at the origin avoiding any bound state.
Therefore, massive gravitons with angular momentum $l>1$ can not be found on the brane, whereas for $l = 0$ resonant states (quasi-localized gravitational massive states highly coupled to the brane \cite{Csaki1, Csaki2}) are possible. It is expected since we have a degenerated massless spectra in Eq. (\ref{Sturm-Liouville-Eq}) and the four dimensional graviton being identified with the $s-$waves ($l=0$). It is important to note that Neumann boundary conditions are satisfied at the brane core position, whereas for $\rho = 0$, $\psi_0^{\prime} = \infty$ due the analogue quantum potential. Consequences of this interaction may be detected phenomenologically as a correction to the Newtonian potential \cite{Gherghetta:2000qi, Csaki1, Csaki2}.

\section{Conclusions and perspectives}
\label{Conclusions and perspectives}

We studied the geometrical and physical properties of a six-dimensional thin-string braneworld extension considering a square dependence on radial extra coordinate on the angular metric component. We proposed a subtle warp function that agrees with the thin string-like model far from the brane and that yields near brane corrections. The geometry possesses a $Z_{2}$ symmetry about the origin and the curvature is well-behaved. Although we have not concerned with a specific physical model, the source of this geometry satisfies the weak energy condition. The energy density has a maximum displaced of the origin and a non-null thickness. 

We also performed the gravity localization on this scenario. The massless mode is localized and, differently from thin-string models, also has a maximum displaced from the origin, which for great values of the cosmological constant, tends to coincide with the energy density one. This shift of the core is due to the conical behavior near the brane. Moreover, for large values of the cosmological constant, the model is reduced to a thin-string model slightly apart from origin. More attention was given to massive modes. We firstly analysed the graviton massive modes equation for its asymptotic regimes and showed that near brane solutions are expressed in terms of Bessel functions of first kind for angular momentum values $l \neq 1$. Far from the brane, we recover Gherghetta-Shaposhnikov thin string model without degenerate states. Numerical solutions of the equation for the massive modes revealed that, in contrast to the thin string model, the massive solutions have considerable amplitude near the brane indicating the possibility gravitational massive states interacting with the defect. 

Finally, we studied the analogue quantum potential. From this formalism we prove the massless mode, which must to satisfy a Sch\"odinger-like equation for $m = 0$. We also conclude that massive states may be found as resonant states near the brane only for $l = 0$ which refers to four dimensional gravitons. As perspectives, these massive states have to be achieved from this potential by suitable numerical methods and be used to predict corrections to the Newtonian potential.

\section{Acknowledgments}

The authors are grateful to Brazilian agencies CNPq, CAPES and FUNCAP for financial support.

\end{document}